\documentclass[a4paper]{jpconf}
\usepackage[T1]{fontenc}
\usepackage{amsmath}
\usepackage{amssymb}
\usepackage{graphicx}
\usepackage{color}

\newcommand{\beq}{\begin{equation}}
\newcommand{\eeq}{\end{equation}}
\newcommand{\la}{\langle}
\newcommand{\ra}{\rangle}
\newcommand{\f}[2]{\frac{#1}{#2}}
\newcommand{\pf}[2]{\frac{\partial #1}{\partial #2}}
\newcommand{\mZ}{\mathcal{Z}}
\newcommand{\mD}{\mathcal{D}}
\newcommand{\mU}{\mathcal{U}}
\newcommand{\tn}[1]{\textnormal{#1}}

\newcommand{\qpsi}[2]{\hat{\psi}_{#1}(#2)}
\newcommand{\qpsidagger}[2]{\hat{\psi}_{#1}^{\dagger}(#2)}

\renewcommand{\eqref}[1]{Eq.~(\ref{Eq:#1})}
\newcommand{\figref}[1]{Fig.~\ref{Fig:#1}}

\begin{document}
\title{A complex Langevin approach to ultracold fermions}
\author{Lukas Rammelm\"uller$^1$, Joaqu\'in E. Drut$^2$, Jens Braun$^{1,3}$}
\address{$^1$ Institut f\"ur Kernphysik (Theoriezentrum), TU Darmstadt, D-64289 Darmstadt, Germany.\\$^2$ Department of Physics and Astronomy, University of North Carolina, Chapel Hill, NC, 27599, USA., \\$^3$ ExtreMe Matter Institute EMMI, GSI, Planckstrasse 1, D-64291 Darmstadt, Germany.}
\ead{lrammelmueller@theorie.ikp.physik.tu-darmstadt.de}

\begin{abstract}
The theoretical treatment of Fermi systems consisting of particles with unequal masses is challenging. Even in one spatial dimension analytic solutions are limited to special configurations and numerical progress with Monte Carlo simulations is hindered by the sign-problem. To circumvent this issue, we exploit the Complex Langevin approach and study one-dimensional mass-imbalanced two-component Fermi gases with attractive and repulsive interactions. We find perfect agreement with results obtained by other methods in a range of parameter space. Promisingly, our approach is not limited to the specific	model presented here and can easily be extended to finite spin polarization and, most notably, can also be applied in higher dimensions.
\end{abstract}

\section{Motivation}

In recent years, tremendous effort was put forward to investigate the physics of ultracold Fermi gases, which led to the observation of a rich variety of phenomena ranging from BCS superfluidity to Bose-Einstein condensation~\cite{Bloch2008, Giorgini2008}. Experimentally, this was made possible by elaborate trapping and cooling techniques as well as by exploiting Feshbach resonances to influence the interaction between particles. Moreover, it became possible to realize experimental setups for low-dimensional quantum gases such that observables in two- and even one-dimensional (1D) systems are now accessible.

This variety of experimentally accessible systems calls for a corresponding theoretical investigation. Unfortunately, the resulting many-body problem is extremely challenging to solve and exact analytical solutions are either unavailable or are limited to very special cases, such as specific models in 1D. Thus, theoretical progress is mostly achieved by numerical methods. Among the most successful methods, in particular for systems beyond the few-body regime, are Quantum Monte Carlo (QMC) approaches. A major drawback, however, is given by the infamous sign-problem, for which a general solution is highly unlikely \cite{Troyer2005}.

In this contribution we seek to address this problem for non-relativistic fermions in the ground state by exploiting the idea of complex stochastic quantization~\cite{Parisi1981}. Although the approach has been known for some time, its rigorous mathematical foundation was unclear at first and applications remained scarce. With revived interest in recent years in the lattice QCD community (see e.g. Ref.~\cite{Aarts2009}), the limitations of this approach have been formulated more rigorously and methods have been developed to mitigate them. Motivated by these advances, we present an approach to ultracold Fermi gases that relies on the Complex Langevin (CL) algorithm. This allows us to extract results in a fully non-perturbative fashion as we demonstrate here for mass-imbalanced fermions in one spatial dimension with attractive and repulsive interactions. The approach is not restricted to 1D calculations but can be applied to systems of any dimension, see also Ref.~\cite{Loheac2017b} for a first status report on the application of this CL approach to the three-dimensional unitary spin-imbalanced Fermi gas.

\section{Path integral for fermions and the sign problem}

In the realm of ultracold dilute Fermi gases, the physics is dominated by s-wave scattering, i.e. contact interaction between particles. In such systems, the resulting Hamilton operator for a two-component system without an external potential is of the form
\beq
	\hat{H} = -\sum_{s=\uparrow,\downarrow}{\sum_{{\bf x}}{\qpsidagger{s}{{\bf x}}\ \f{\hbar^2}{2m_s}\vec{\nabla}^{\,2}\ \qpsi{s}{{\bf x}}}}\ + \
	g\sum_{{\bf x}}{\qpsidagger{\uparrow}{{\bf x}}\,\qpsi{\uparrow}{{\bf x}}\,\qpsidagger{\downarrow}{{\bf x}}\,\qpsi{\downarrow}{{\bf x}}}\,,
	\label{Eq:model}
\eeq
where ${\bf x} = (x_1, \cdots , x_d)$ denotes the d-dimensional vector of spatial coordinates (which we will discretize) and $g$ implies attractive (repulsive) interaction for $g <0$ ($g>0$). As usual, $\qpsi{s}{{\bf x}}$ and $\qpsidagger{s}{{\bf x}}$ represent annihilation and creation operators for a particle of flavor $s$ at the coordinate ${\bf x}$, respectively. While the first term, corresponding to the kinetic energy, is rather straightforward to deal with, the second term, i.e. the interaction part of the Hamiltonian, renders the problem difficult to solve due to its two-body nature. Although exact diagonalization is in principle possible, such a solution scales very poorly with the system size, such that other methods are needed to deal with systems beyond the few-body level.

To this end, we employ a projective method to filter out the excited states and obtain the ground state starting from an initial trial state $|\psi_T\ra$, which we take to be a Slater determinant:
\beq
	\mZ_\beta = \la\psi_T|\,e^{-\beta\hat{H}}\,|\psi_T\ra \equiv \la\psi_T|\,\hat{\mU}_\beta\,|\psi_T\ra,
	\label{Eq:projection}
\eeq
The true ground state is reached in the limit of large imaginary time $\beta$, i.e. $\beta \rightarrow \infty$. In order to compute the evolution operator $\hat{\mU}_\beta$, we first discretize imaginary time with a Trotter-Suzuki decomposition and subsequently perform a Hubbard-Stratonovich (HS) transformation to reduce the two-body operator in \eqref{model} into a sum over one-body operators coupled to an external (auxiliary) field \cite{Blankenbecler1981, Drut2013}. All contributions from this high-dimensional sum can be collected into a path integral which yields the partition function $\mZ$ and is of the form
\begin{equation}
	\mZ\ =\ \int{\mD \sigma\ \det{U_\sigma^{(\uparrow)}}\  \det{U_\sigma^{(\downarrow)}}}\ \equiv\ \int{\mD \sigma\ e^{-S[\sigma]}},
	\label{Eq:partition_function}
\end{equation}
where $U_{\sigma}^{(s)}$ denotes the imaginary time evolution operator of the spin flavor $s$ in single-particle representation. It can easily be appreciated from the above expression that the integrand is in general not positive semidefinite or in other words, the action $S_\sigma$ is in general a complex quantity. The complex nature of the action restricts any probabilistic approach (such as importance sampling with the Metropolis algorithm) to cases where its imaginary part vanishes, i.e. to systems with equally populated spin species of equal mass and attractive pairwise interaction. In any other case such an algorithm exhibits a sign-problem and thus, the computational efforts grow exponentially with the system size. This is this reason that also restricts simulations of the repulsive Hubbard model to the half-filled case (see, e.g., Ref.~\cite{RepulsiveHubbard}) as well as lattice QCD to zero baryon chemical potential (see, e.g., Ref.~\cite{AartsLectures, Philipsen2013}). In the following, we introduce an approach to surmount this issue by means of stochastic quantization.

\section{Complex Langevin approach}
The main idea of stochastic quantization is to use a random process, governed by the Langevin equation, to generate a probability measure~$\sim {\rm e}^{-S}$ that corresponds to the one in the Euclidean path integral in \eqref{partition_function}. More specifically, one introduces a fictitious Langevin time $ t_{CL} $ and evolves the auxiliary field $ \sigma $ according to
\beq
	\dot{\sigma}(t)\ =\ -\pf{S[\sigma(t)]}{\sigma(t)}\ +\ \eta(t),
	\label{Eq:real_langevin}
\eeq
where the dot represents a derivative with respect to $t_{CL}$ and $\eta(t)$ is a random noise term with $\la \eta(t) \ra = 0$ and
$\la\eta(t)\,\eta(t')\ra = 2\delta_{t,t'}$. Note that there is one component of the auxiliary field for every spacetime lattice site. In the limit $t_{CL} \rightarrow \infty $, the distribution of $\sigma$ (ideally) becomes stationary and follows $\sim {\rm e}^{-S[\sigma]}$.

While this elegant approach works well for real $S$, complex $S$ requires complexifying the auxiliary field and more care. Indeed, the Langevin equation acquires an imaginary part and, after discretizing $t_{CL}$, the evolution equations become
\beq
	\begin{split}
			\delta{\sigma_R}\ &=\ -\tn{Re}\left[\pf{S[\sigma]}{\sigma}\right]h_t\  + \ \eta \sqrt{h_t}\,,\\
			\delta{\sigma_I}\ &=\ -\tn{Im}\left[\pf{S[\sigma]}{\sigma}\right]h_t,
	\end{split}
	\label{Eq:complex_langevin}
\eeq
where $h_t$ serves as the integration time step. Although similar in structure to the real Langevin equation, two main problems arise: first, the random process may not converge and furthermore, even if the a stationary distribution is approached there is no guarantee that the results are correct and therefore further checks are in general required. The former issue is of a numerical nature and can be cured by choosing the integration step $h_t$ to be adaptive \cite{Aarts2010}. The possibly faulty convergence, on the other hand, poses a much more serious problem and was traced back to the occurrence of singularities in the drift term $\sim \pf{S[\sigma]}{\sigma}$ in \eqref{complex_langevin}, which correspond to zeroes in the fermion determinant in our approach. While it is difficult to gain insight into the pole structure of this complicated quantity before performing the actual simulation, it is possible to analyze its behavior {\it a posteriori} as pointed out in Ref.~\cite{Aarts2017}.

The HS transform chosen for this work, which is typical in hybrid Monte Carlo (HMC) simulations, is proportional to $\sin(\sigma)$ and thus bounded in the real part but unbounded in the imaginary direction. In order to avoid singularities associated with an ``unboundedly escaping" $\sigma$ field, we amend the CL equations with an extra term $\sim \xi$ which we call the ``regulator'' \cite{Loheac2017}:
\beq
  \begin{split}
		\delta{\sigma}_R\ &=\ -\tn{Re}\left[\pf{S[\sigma]}{\sigma}\right]h_t\ -\ 2\xi\sigma_{\tn{R}}h_t + \ \eta \sqrt{h_t}\,,\\
		\delta{\sigma}_I\ &=\ -\tn{Im}\left[\pf{S[\sigma]}{\sigma}\right]h_t\ -\ 2\xi\sigma_{\tn{I}}h_t\,.
  \end{split}
\eeq
The extra term can be thought of as a damping force that keeps the magnitude of the HS field from ``escaping" too far into the complex plane. Naturally this introduces a systematic error which has to be studied carefully. By performing runs at different small but finite values of~$\xi$ and then extrapolate to the limit $\xi\rightarrow 0$ one can eventually extract the results for physical observables~\cite{Loheac2017, Rammelmueller2017b}.

\section{Case study: 1D fermions with unequal masses}
Equipped with a method that is able to deal with systems that are subject to a sign problem in case of conventional QMC methods, we now turn to a first application: the ground state of 1D fermions with unequal masses. We consider Fermi mixtures of two equally populated species but finite mass difference, according to the model in~\eqref{model}. Note that even in 1D there is no general analytic solution to this problem and so far all attempts are restricted to specific mass/particle configurations \cite{Olshanii2015, Harshman2017}.

For all calculations in the following, we put a fixed number of $N = N_\uparrow + N_\downarrow$ fermions on a 1D lattice of $N_x = 40$ sites and enforce periodic boundary conditions to minimize finite-size effects. While our approach is by no means limited to such a configuration, we found it sufficient to gain a first insight and provide a benchmark for ground-state calculations in a non-relativistic setting with the CL method. In order to extrapolate to the ground state we performed simulations at large imaginary times up to $\sim 1500$ temporal steps with a step size of $d\tau=0.05$ and subsequently averaged over several runs with varying temporal extent, which was found sufficient in previous studies with the HMC method \cite{Rammelmueller2017a}. Furthermore, all results reflect averages over $5000$ de-correlated samples originating from a number of different random seeds, defining the initial conditions of the CL equations. In the following, we use the convention~$\hbar = c = m_0 = 1$, where $m_0$ is the average mass across the two species. We quantify the mass-imbalance by the dimensionless mass-asymmetry parameter
\beq
	\bar m = \f{m_\uparrow - m_\downarrow}{m_\uparrow + m_\downarrow}.
	\label{Eq:mbar}
\eeq
In order to present all results as dimensionless quantities, we rescale energies with the non-interacting, mass-balanced energy in the continuum:
\beq
	E_{FG} = \f{1}{3}N\varepsilon_F = \f{1}{6} N {k_F^2}
	\label{Eq:efg}
\eeq
where $k_F = \f{\pi}{2}n$ is the Fermi wavevector. Additionally, we use the (constant) fermion density $n$ to define the dimensionless coupling $\gamma = g/n$, as it is typical in the literature of 1D systems~\cite{Barth2011}.

\begin{figure}
	\includegraphics[width=0.495\linewidth]{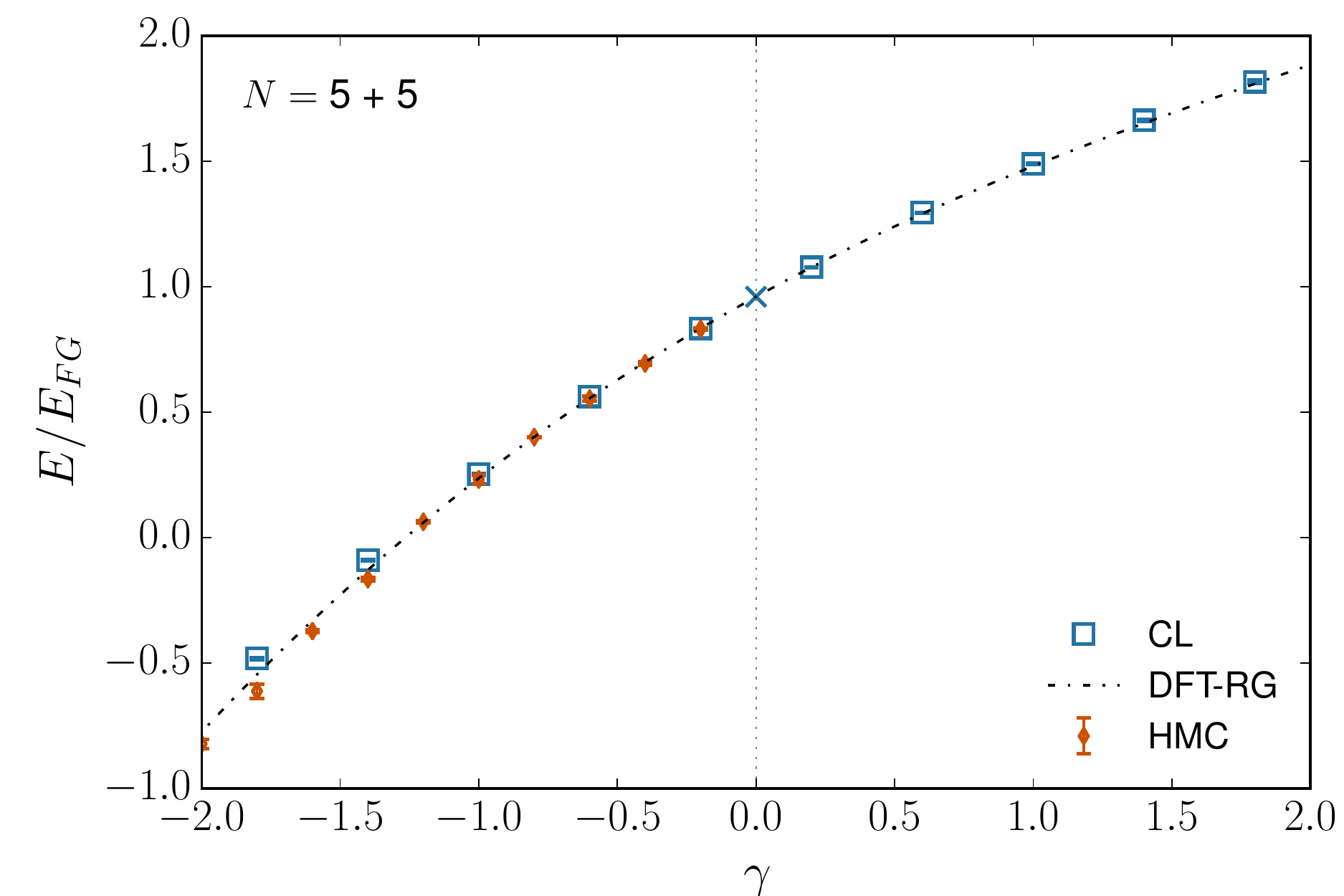}
	\includegraphics[width=0.495\linewidth]{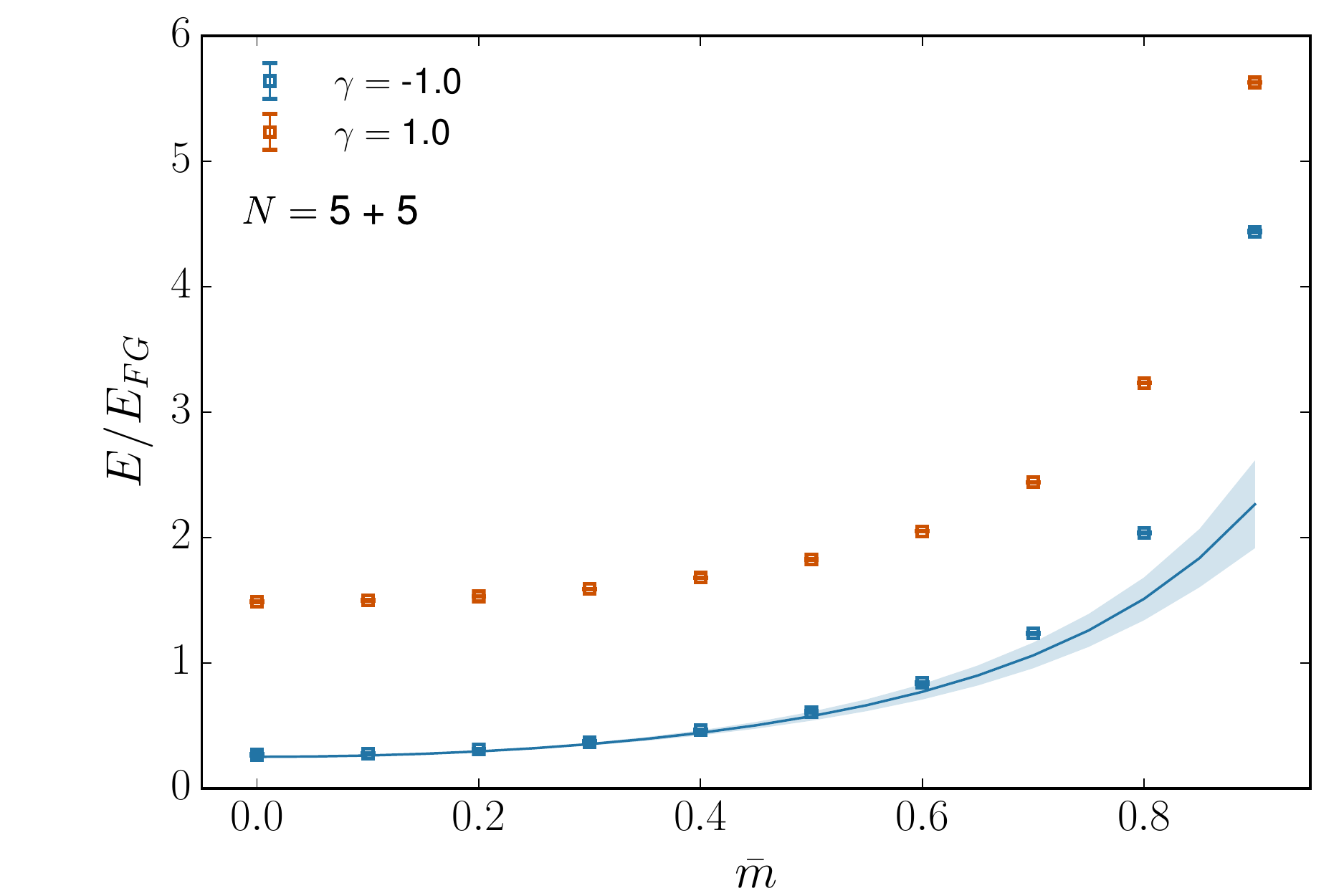}
	\caption{Ground-state energy (normalized with the non-interacting value $E_{FG}$) for systems of $N = 5+5$ particles on a lattice with $N_x = 40$ sites. Left:  energies as a function of the dimensionless coupling $\gamma$. We show CL results (blue squares) alongside with HMC values for attractive couplings (red diamonds) \cite{Rammelmueller2015} and DFT-RG results for attractive and repulsive interactions (dashed-dotted line) \cite{Kemler2017, Kemler2017b}. Right: energies as a function of the mass asymmetry parameter $\bar m$ for couplings of $\gamma = 1.0$ and $\gamma = -1.0$. Symbols depict CL results whereas solid lines represent analytically continued iHMC results where the shaded area is the $95 \%$ confidence level associated with the corresponding fits of the data.}
	\label{Fig:compare}
\end{figure}

\subsection{Benchmark study for balanced systems}
We first consider systems with $\bar m = 0$, i.e. mass-balanced mixtures. In this case, we are able to compare with a variety of other methods as shown in the left panel of~\figref{compare}. In the attractive regime ($\gamma < 0$) our CL values agree very well with results obtained from HMC simulations, which have been shown to agree with the solutions based on the Bethe ansatz~\cite{Rammelmueller2015}. Additionally, we show results from a renormalization group approach to density functional theory (DFT-RG) and observe perfect agreement for $|\gamma| \lesssim 2.0$, which is the regime where the DFT-RG approach is currently expected to yield reliable results \cite{Kemler2017}.

\subsection{Feasibility study: systems with arbitrary mass asymmetry}
By moving to finite $\bar m$, the possibilities to cross-check our result are much more limited as this regime is harder to access, especially at large imbalances. One method that is able to provide results up to medium mass imbalances is the so-called iHMC approach: by taking the mass-asymmetry to be imaginary, one obtains a positive-semidefinite measure in~\eqref{partition_function}, ultimately necessary for the HMC approach. The price to pay is an analytic continuation from imaginary $\bar m$ to the real axis, which requires very precise expectation values in order to push the results to large $\bar m$. The method was successfully applied to the unitary Fermi gas \cite{Braun2015} as well as 1D fermions with mass-imbalance in Ref.~\cite{Rammelmueller2017b}, to which we compare the CL results in the right panel of~\figref{compare}. Remarkably, the CL and iHMC results are practically indistinguishable up to $\bar m \sim 0.5$, which is where artefacts associated with the analytic continuation in the iHMC algorithm start to become noticeable (indicated by the large uncertainty bands). In contrast, the CL results remain smooth in this large-$\bar{m}$ regime and associated error bars stay roughly of the same magnitude as for~$\bar{m}\lesssim 0.5$.

\subsection{Full equation of state for 1D fermions with unequal masses}
\begin{figure}
	\centering
	\includegraphics[width=0.7\linewidth]{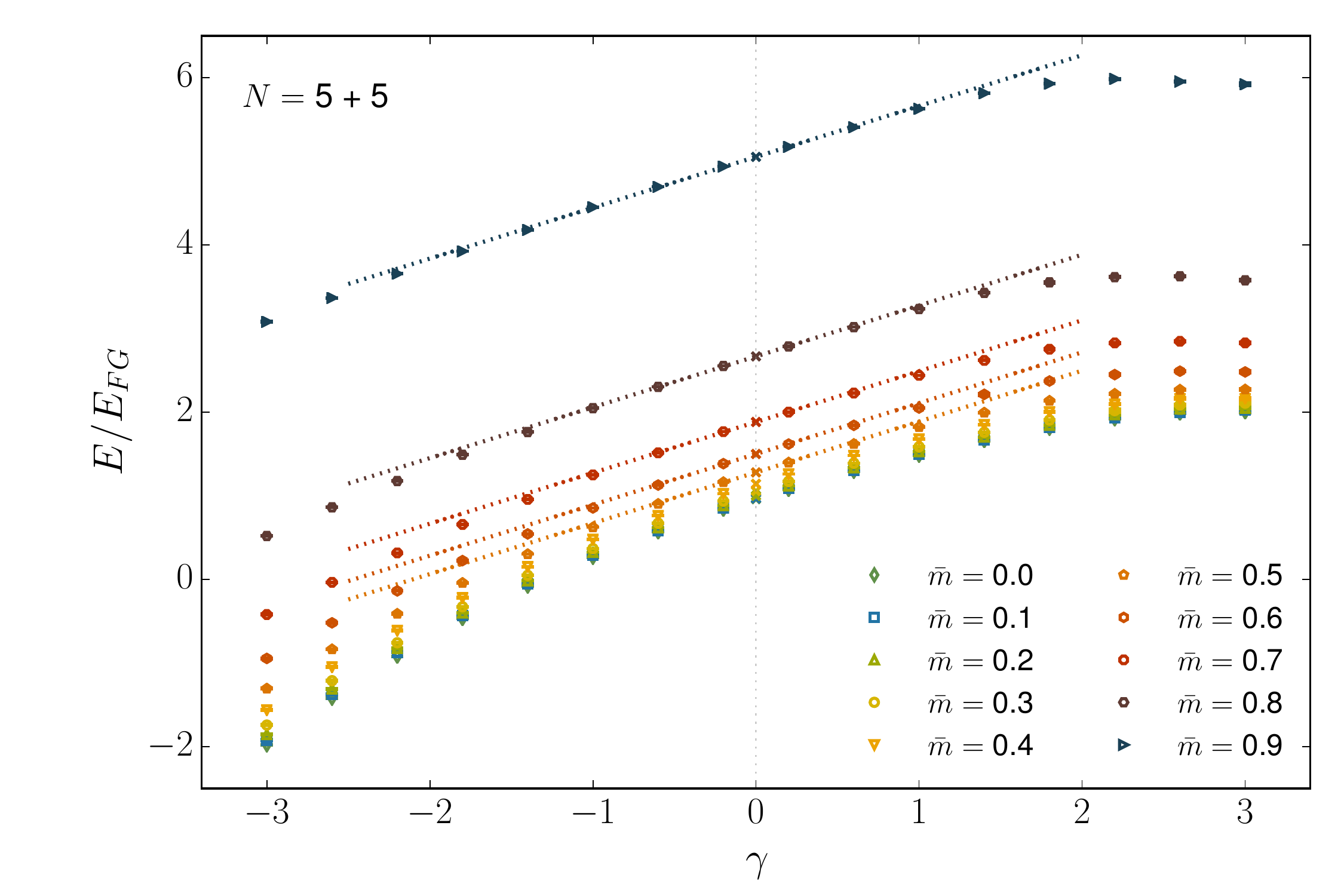}
	\caption{Ground-state energy as a function of the coupling $\gamma$ for $N = 5+5$ fermions on a lattice of $N_x = 40$ sites	for $\bar m = 0.0, 0.1, \dots , 0.9$. The symbols show values from the CL approach together	with values obtained from first-order lattice perturbation theory (dotted lines).}
	\label{Fig:full_EOS}
\end{figure}

Backed up with the excellent agreement with a variety of methods, we now present the full equation of state for a large array of mass imbalances as a function of the dimensionless coupling~$\gamma$ in~\figref{full_EOS}. The wide range of parameters studied here underscores the versatility of our CL approach as it includes regimes relevant to (future) experimental setups where quantitative results are scarce. Such experiments include Fermi mixtures of $^{6}$Li and $^{40}$Ka as well as $^{40}$Ka and $^{161}$Dy, which correspond to $\bar m\,=\,0.74$ and $\bar m\,=\,0.6$ in our convention, respectively.

It is necessary to mention, however, that results for repulsive interactions have to be taken with caution. For repulsive couplings we observe outliers in the distributions of the studied observables and the number of these outliers even increases with increasing coupling strength. This observation can be traced back to two different effects: First, there is a signal-to-noise issue associated with methods that use the fermion determinant as a probability measure which results in larger uncertainties than the ones associated with an assumed Gaussian distribution \cite{Hao2016}. Second, there are implications associated with the occurrence of zeroes in the fermion determinant within the CL framework \cite{Aarts2017}. While the first issue is solvable by numerical means, albeit not implemented in this work, the second is of a conceptual nature, as mentioned above. At the present stage, however, it is difficult to separate one from the other and it is subject of ongoing investigations to resolve the origin of the observed behavior of the distributions of observables at strong repulsive couplings.

\section{Conclusion \& Outlook}
In this contribution we have used a modified version of an algorithm known from the lattice QCD community and applied it to non-relativistic systems, allowing us to investigate areas of the parameter space in non-relativistic systems that previously have been prohibitive. As a cross-check we have performed comparisons with various methods and found remarkable agreement with our CL results. Furthermore, we computed the full equation of state for spin-balanced but mass-imbalanced fermions in one dimension, a problem that is beyond the capabilities of previously existing stochastic methods.

Although the CL approach has been known for many years and has been applied frequently in relativistic studies, it is still a method under construction, at least in non-relativistic settings. Its main benefit is the extendibility to systems at finite temperature, polarized systems as well as arbitrary dimension. Additionally, it is possible, in a straightforward manner,  to compute more sophisticated observables such as density matrices and correlation functions, which enables the investigation of inhomogeneous phases, such as the Fulde-Ferell-Larkin-Ovchinnikov phase. The application of our approach to such phenomena is currently under investigation.

\ack
This work was supported by HIC for FAIR within the LOEWE program of the State of Hesse and the U.S. National Science
Foundation under Grant No. PHY1452635 (Computational Physics Program). Numerical calculations have partially been performed at the LOEWE-CSC Frankfurt.



\end{document}